\def\year{2022}\relax
\documentclass[letterpaper]{article} % DO NOT CHANGE THIS
\usepackage{aaai22}  % DO NOT CHANGE THIS
\usepackage{times}  % DO NOT CHANGE THIS
\usepackage{helvet}  % DO NOT CHANGE THIS
\usepackage{courier}  % DO NOT CHANGE THIS
\usepackage[hyphens]{url}  % DO NOT CHANGE THIS
\usepackage{graphicx} % DO NOT CHANGE THIS
\usepackage{newfloat}
\usepackage{listings}
\usepackage{amssymb}
\usepackage{fontawesome5} 
\usepackage[most]{tcolorbox}
\usepackage{lipsum} 
\usepackage{natbib}  % DO NOT CHANGE THIS AND DO NOT ADD ANY OPTIONS TO IT
\usepackage{caption} % DO NOT CHANGE THIS AND DO NOT ADD ANY OPTIONS TO IT
\DeclareCaptionStyle{ruled}{labelfont=normalfont,labelsep=colon,strut=off} % DO NOT CHANGE THIS
\usepackage{algorithm}
\usepackage{algorithmic}
\usepackage{amsmath}
\usepackage{tabularray}
\usepackage{newfloat}
\usepackage{listings}
\floatstyle{ruled}
\newfloat{listing}{tb}{lst}{}
\floatname{listing}{Listing}

\title{HCAG: Hierarchical Abstraction and Retrieval-Augmented Generation on Theoretical Repositories with LLMs}
\author {
    Yusen Wu,
    Xiaotie Deng
}
\affiliations {
}

\usepackage{bibentry}
\begin{document}

\maketitle

\begin{abstract}
  Existing Retrieval-Augmented Generation (RAG) methods for code struggle to capture the high-level architectural patterns and cross-file dependencies inherent in complex, theory-driven codebases, such as those in algorithmic game theory (AGT), leading to a persistent semantic and structural gap between abstract concepts and executable implementations. To address this challenge, we propose \emph{Hierarchical Code/Architecture-guided Agent Generation} (HCAG), a framework that reformulates repository-level code generation as a structured, planning-oriented process over hierarchical knowledge. HCAG adopts a two-phase design: an offline hierarchical abstraction phase that recursively parses code repositories and aligned theoretical texts to construct a multi-resolution semantic knowledge base explicitly linking theory, architecture, and implementation; and an online hierarchical retrieval and scaffolded generation phase that performs top-down, level-wise retrieval to guide LLMs in an architecture-then-module generation paradigm. To further improve robustness and consistency, HCAG integrates a multi-agent discussion inspired by cooperative game. We provide a theoretical analysis showing that hierarchical abstraction with adaptive node compression achieves cost-optimality compared to flat and iterative RAG baselines. Extensive experiments on diverse game-theoretic system generation tasks demonstrate that HCAG substantially outperforms representative repository-level methods in code quality, architectural coherence, and requirement pass rate. In addition, HCAG produces a large-scale, aligned theory–implementation dataset that effectively enhances domain-specific LLMs through post-training. Although demanstrated in AGT, HCAG paradigm also offers a general blueprint for mining, reusing, and generating complex systems from structured codebases in other domains.
\end{abstract}

\section{Introduction}

Traditional research domains has evolved over decades, accumulating a rich body of theoretical knowledge and practical implementations. They have been applied across diverse domains such as economic mechanism design, policy optimization, multi-agent systems, and industrial automation\cite{agt-usecase1,agt-usecase2,agt-usecase3,agt-usecase4}. Much of this applied knowledge is encapsulated in code repositories and documentation. However, these repositories often present significant challenges for automated understanding and reuse\cite{agt-complex1,agt-complex2,agt-complex3}. They are typically large, sparsely or inconsistently annotated, and contain domain-specific noise. Crucially, the complete logic of a game-theoretic model or system is rarely contained within a single file; instead, it is distributed across a hierarchical directory structure involving multiple files and modules (e.g., separate definitions for buyer and seller types in an auction). This structural complexity makes it difficult for both humans and machines to retrieve and synthesize the relevant architectural knowledge needed for new tasks with current approach mainly rely on mannual design\cite{agt-current1, agt-current2}. Here we use Algorithmic Game Theory (AGT) as a study example, but we also demonstrate the challenge is not limited to it.

Retrieval-Augmented Generation (RAG) has emerged as a promising paradigm to enhance large language models (LLMs) with external knowledge. However, traditional RAG methods—whether using machine learning \cite{base1,base2,base3,base4,base5,base6,base7,base8} or LLM enhancement \cite{llm1,llm2,llm3,llm4,llm5,llm6,llm7,llm8,llm9,llm10,llm11}—typically operate on flat, unstructured text or code snippets. They struggle to capture and reason over the high-level architectural patterns and cross-file dependencies inherent in complex codebases like those in AGT. To address this, some repository-level retrieval methods with LLM reasoning in the retrieval process \cite{coderag,arks,repocoder,knowledge-graph,semantic} have been introduced. However, these approaches, which consume multiple LLM inference queries, are often computationally inefficient, especially when task complexity increases. When tasked with generating a novel system (e.g., a new auction simulation) from a natural language description, LLMs face a dual challenge: a \textit{semantic gap} between the abstract problem description and the concrete implementation logic, and a \textit{structural gap} in piecing together a coherent, executable system from fragmented references. While recent benchmarks like TMGBench \cite{tmgbench} and GLEE \cite{glee} have advanced strategic reasoning evaluation, and methods like Consensus Game \cite{consensus} have infused game-theoretic concepts into decoding, the problem of \textit{systematic generation} of verifiable implementations from theoretical concepts remains largely unaddressed.

To bridge these gaps, we introduce the \textbf{Hierarchical Code/Architecture-guided Agent Generation (HCAG)} framework. HCAG reformulates repository-level code generation into a structured retrieval and planning process. Its core innovation is a two-phase methodology: (1) an offline \textit{hierarchical abstraction phase} where an LLM recursively parses a codebase to build a multi-layered, symbolic summary capturing functional semantics and architecture across granularities; and (2) an online \textit{hierarchical retrieval and generation phase}, where given a new task the framework performs a top-down, level-wise search through this abstracted structure to retrieve the most relevant architectural skeletons and implementation snippets. This process guides the LLM in a scaffold-and-fill manner, first establishing a coherent system architecture and then instantiating in module-level details. In addition, HCAG incorporates a multi-agent discussion mechanism, inspired by cooperative game \cite{consensus}, to iteratively debate and refine the generated design, thereby improving its consistency and robustness.

The primary contributions of this work are threefold: We propose a novel framework that moves beyond flat text retrieval by creating and leveraging a hierarchical, semantic summary of code repositories. This enables LLMs to retrieve and reason over architectural knowledge, leading to the generation of more coherent and executable systems for complex domains like AGT. HCAG jointly processes theoretical texts (e.g., papers) and their corresponding implementations, creating an aligned knowledge base. This alignment allows the framework to better map abstract theoretical requirements (e.g., "design a repeated game with a trigger strategy") to concrete, executable code structures. Finally while demonstrated in the AGT domain, HCAG methodology is general. It provides a blueprint for handling complex, structured codebases in other fields with similar challenges (e.g., online learning, reinforcement learning as shwon in appendix). The hierarchical summaries also constitute the first large-scale, structured dataset for code-function alignment, offering a valuable resource for training domain-specific LLMs.

\section{Related Work}

Our work builds upon two critical research thrusts: (1) leveraging LLMs to understand, apply, and generate game-theoretic constructs, and (2) advancing repository-level code retrieval and generation. HCAG synthesizes insights from these areas to enable the systematic \textit{mining, rediscovery, and operationalization} of knowledge embedded in theoretical codebases.

\textbf{Mining and Rediscovering Game Theory with LLMs.} A primary challenge in applying LLMs to specialized domains like AGT is the scarcity of dedicated training data. Recent research has tackled this issue by probing and augmenting LLMs' strategic reasoning. One strand constructs benchmarks to evaluate LLMs' grasp of concepts like Nash equilibrium and theory of mind within structured environments \cite{tmgbench,glee}. Other studies leverage LLMs as agents within strategic simulations to analyze emergent behaviors \cite{alympics}. Complementary work seeks to instill game-theoretic reasoning into the decoding process, e.g., by framing generation as an equilibrium-finding problem \cite{consensus}. While valuable, these approaches focus on how an LLM \textit{plays} a given game, not on how it can \textit{discover} and \textit{reconstitute} the game's underlying computational model from prior work and do not address the knowledge acquisition problem: enabling an LLM to navigate implicit knowledge in existing AGT implementations to synthesize new systems. 

\textbf{Repository-Level Code Retrieval and Generation (RAG).} Enabling LLMs to generate complex, multi-file systems requires moving beyond document-level retrieval. Recent repository-level RAG methods aim to provide richer contextual grounding. Some approaches leverage structured representations such as code knowledge graphs \cite{coderag}, but their reliance on static analysis makes them brittle in domains with unconventional or highly dynamic code patterns. Other methods improve retrieval through query rewriting \cite{semantic} or iterative, feedback-driven context expansion \cite{arks}. However, these techniques often operate in a \textit{flat} or \textit{reactive} manner, retrieving isolated snippets or incrementally expanding context without a proactive understanding of the repository’s architectural blueprint. This limitation is particularly pronounced for theoretical codebases where core design logic emerges from interactions distributed across a hierarchical structure.

\section{Preliminaries}

We first formalize the traditional RAG paradigm for code generation, then analyze the limitations of representative repository-level methods in complex, theory-aligned domains like AGT.

\subsection{Traditional RAG for Code Generation}
The canonical RAG pipeline for code generation retrieves relevant snippets $C = {c_1, c_2, ..., c_k}$ from a knowledge base given a user query $Q_{user}$, using a retrieval function $\mathcal{R}(Q_{user}) \rightarrow C$ typically based on dense vector similarity. These contexts are concatenated with $Q_{user}$ to form an augmented prompt $P_{aug} = [Q_{user}; C]$ for an LLM. This paradigm exhibits critical shortcomings for generating complete systems: flat retrieval granularity, where retrieved contexts are isolated, non-hierarchical blocks lacking architectural roles or inter-dependency information; semantic--structural gap, where semantic similarity alone poorly supports the mapping from high-level task descriptions to low-level implementation details; and context window inefficiency, where retrieving many fragments to cover system-wide logic can exceed the LLM’s context window, while large chunks introduce noise. These limitations are acute in AGT where a single conceptual mechanism is realized through coordinated logic distributed across multiple modules.

\subsection{Advanced Repository-Level RAG Methods}
We analyze why representative repository-level methods remain insufficient for effectively mining and utilizing knowledge from complex codebases.

\textbf{Static Graph-Based Methods} (e.g., CodeRAG \cite{coderag}, PKG-based \cite{knowledge-graph}) construct explicit knowledge graphs capturing entities and relations. Their key strength lies in explicit structure awareness, but they face a fundamental \textit{reliability--flexibility} tradeoff. Graph construction relies heavily on static program analysis, which often fails for dynamically-typed languages, templated code, or sparsely documented repositories—common in AGT. The resulting "hard" graph is brittle: missing or erroneous edges degrade retrieval quality. Moreover, predefined graph schema may not align with task-oriented functional abstractions.

\textbf{Dynamic \& Iterative Retrieval Methods} (e.g., ARKS \cite{arks}, RepoCoder \cite{repocoder}) adopt an iterative generate-then-retrieve loop. ARKS dynamically formulates new queries based on generation feedback, producing a loosely structured “knowledge soup” while RepoCoder iteratively retrieves context based on previously generated code to resolve cross-file dependencies. Their advantage is adaptability to the immediate generation context. The core drawback is their \textit{reactive and local} nature. They lack a global, proactive understanding of repository architecture. Each retrieval step is a local optimization, which can lead to inefficiency (many iterations), myopia (getting stuck in a suboptimal architectural path), and compounded errors from early mistakes. For generating a complete system from scratch, this reactive paradigm resembles exploring a maze without a map.

\textbf{Query-Augmentation Methods} (e.g., LLM Agents for Search \cite{semantic}) focus on refining the user query \(Q_{user}\) into a more search-effective query \(Q'_{user}\) using an LLM agent informed by repository metadata. This approach directly attacks vocabulary mismatch. However, its effectiveness is inherently bounded by the agent’s often-shallow understanding of the repository’s deep functional structure, gleaned only from surface-level docs like READMEs. It does not fundamentally alter the \textit{retrieval corpus} from being a collection of flat documents. The enhanced query still retrieves from the same unstructured or weakly-structured pool, leaving the core problem of architectural knowledge retrieval unsolved.

The analysis reveals a common theme: existing methods focus on improving the \textit{retrieval mechanism} over a corpus that remains fundamentally \textit{flat} or \textit{rigidly pre-structured}. They do not proactively transform the raw codebase into a multi-resolution, functionally organized knowledge representation that mirrors human design reasoning. This gap is particularly consequential for domains like AGT, where valuable knowledge is encoded in hierarchical composition patterns rather than individual lines of code. Effective mining and reuse require transforming the codebase into a "soft" LLM-generated functional hierarchy—a lightweight, adaptable map facilitating top-down, planning-driven retrieval and generation. This insight directly motivates HCAG's design.

\section{Methodology}

\subsection{Motivation and Design Principles}

The application of Large Language Models (LLMs) to complex system generation, agent planning, and game-theoretic reasoning exposes significant structural limitations in conventional Retrieval-Augmented Generation (RAG) paradigm. First, retrieval operates at a flat granularity, fetching isolated text or code snippets that fail to capture the architectural logic embedded in mature systems. Second, although LLMs can reason about abstract strategic concepts (e.g., Nash equilibrium, bargaining dynamics), they lack an integrated "theory-architecture-implementation" pipeline,  which hinders the effective translation of theoretical insights into executable code. Third, specialized domains like AGT are characterized by data scarcity and high levels of abstraction, making it difficult for models to internalize deep logical structures from limited examples.

Hierarchical Code/Architecture-Guided Generation(HCAG) framework is designed to bridge these three gaps. It transforms theoretical and implementation-level knowledge of domains like game theory into a structured, retrievable, and reusable resource for system generation. HCAG introduces a two-tiered "architecture-then-module" generation mechanism, ensuring that theoretical reasoning is grounded into a verifiable system implementations. This philosophy aligns with efforts like ChessGPT \cite{chessgpt} in seeking to internalize verified experience but extends it by externalizing this experience into an explicit, retrievable knowledge base. In contrast to benchmarks like TMGBench \cite{tmgbench} or GLEE \cite{glee} that focus on evaluating strategic behavior, HCAG emphasizes the closed-loop reuse of structured knowledge for generation. Unlike methods like Consensus Game \cite{consensus} that apply game theory to enhance decoding, HCAG leverages game-theoretic principles to enhance \textit{system architecture generation}. This shift offers two major gains: it provides LLMs with a structured and validated architectural scaffold rather than generating from scratch, and it establishes a direct mapping pathway from abstract theory to executable code modules.

\subsection{Phase I: Codebase Abstraction and Hierarchical Summarization}
The core of HCAG is a proactive, offline process that transforms a raw code repository into a multi-resolution, semantically indexed knowledge base.  This stands in contrast to the reactive and flat retrieval mechanisms employed by standard RAG methods.

\begin{figure}[htbp]
    \centering
    \includegraphics[width=0.50\textwidth]{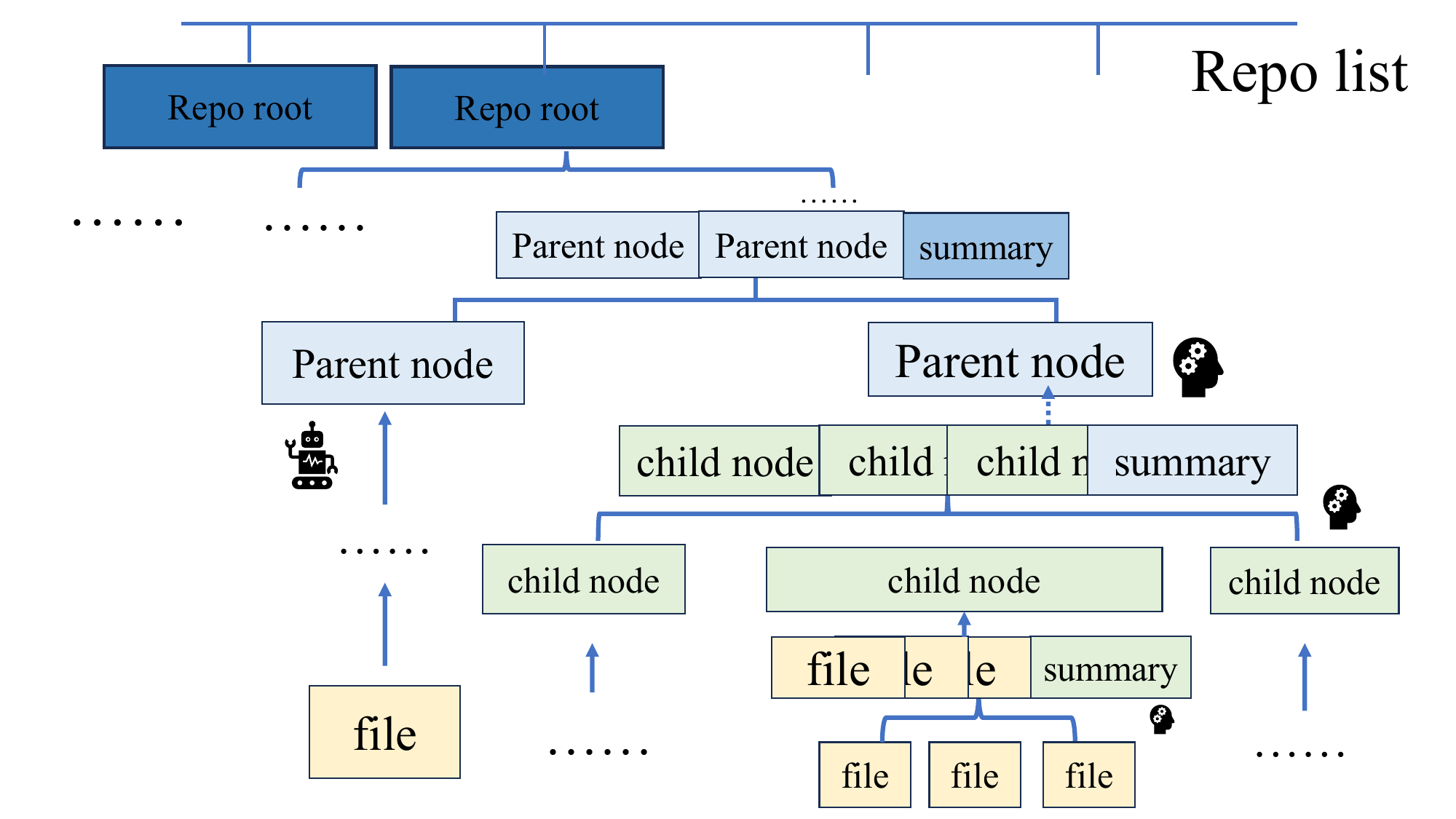}
    \caption{Phase I Pipeline: Recursive Abstraction \& Hierarchical Summarization.
    For each target codebase, HCAG recursively extracts logical summaries (green and yellow blocks) from the underlying textual content (orange blocks). At each level, the original content (index) and its summary are aggregated upward until the root node is reached. For nodes exceeding a configurable depth threshold, compression is applied (indicated by ellipsis): a placeholder label is generated, which can be expanded on-demand during later retrieval when the node is first accessed.}
    \label{graph: phase1}
\end{figure}

\textbf{Recursive Parsing and Summarization.}Given a target code repository, HCAG employs an LLM in a recursive, bottom-up process to construct a hierarchical summary. Starting from leaf files, the LLM generates a semantic summary for each file, capturing its \textit{key functionality, core logic, inputs/outputs, and dependencies}. For directories (non-leaf nodes), the summaries of constituent files or subdirectories are aggregated and presented to the LLM to generate a higher-level summary that characterizes the module's overall purpose and architecture. This process recurses upward until the repository root. The pipeline for this phase is illustrated in \ref{graph: phase1}.

A key design consideration is the handling of deep hierarchies. To balance efficiency and focus, a configurable compression depth parameter \(C\) is introduced. When the directory depth exceeds \(C\) during parsing, detailed summarization for that subtree is deferred, and a placeholder summary (e.g., 'to be detailed') is generated instead along with an index pointing to the underlying content. This creates a "soft" compression, where deep structural details can be expanded on-demand during later retrieval, trading initial abstraction cost for downstream flexibility.

\textbf{Structured Knowledge Base Construction.}The output of this phase is a structured, tree-like knowledge base \(\mathcal{K}\). Each node \(n_i\) in \(\mathcal{K}\) is represented as a tuple:
\[
n_i = (\texttt{path}_i, \texttt{abstract}_i, \{\texttt{submodule}_j\}, \texttt{content\_ref}_i)
\]
where \(\texttt{abstract}_i\) denotes the LLM-generated semantic summary, \(\{\texttt{submodule}_j\}\) is the list of child nodes (empty for leaves), and \(\texttt{content\_ref}_i\) is a pointer to the actual code content (or compressed placeholder). This representation forms a lightweight "soft knowledge graph" that is semantically rich, adaptive while avoiding the brittleness of static analysis-based graphs used in methods like CodeRAG \cite{coderag}.

\subsection{Phase II: Hierarchical Retrieval and Scaffolded Generation}
\begin{figure}[htbp]
    \centering
    \includegraphics[width=0.50\textwidth]{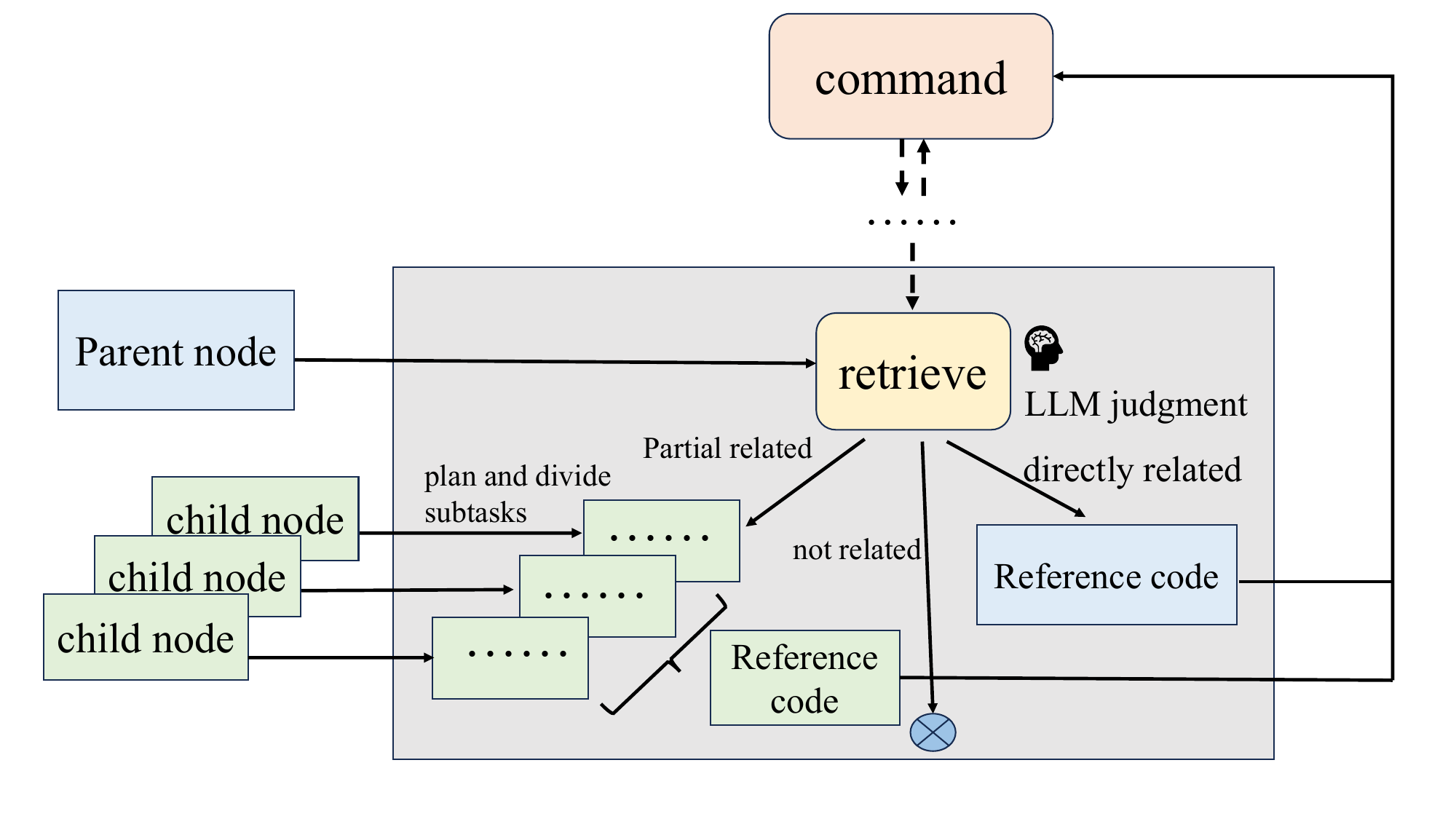}
    \caption{Phase II Pipeline: Hierarchical Retrieval and Scaffolded Generation.
    Given a user task instruction, the LLM evaluates the relevance of each retrieved node with respect to the task. If a node is fully relevant, its content is returned; if it is irrelevant or a leaf node, retrieval along that branch terminates. More commonly, a node is partially relevant, triggering task decomposition over its sub-nodes and recursive retrieval. Encountering a compressed placeholder node initiates on-demand expansion using the Phase~I abstraction procedure to generate structured summary for the subtree. This recursive process continues(gray boxes denote parent nodes, green boxes denote child nodes with identical internal structure) until the structured codebase abstraction has been sufficiently traversed to support generation.}
    \label{graph: phase2}
\end{figure}

Upon receiving a new user query \(Q_{user}\) (e.g., "Implement a simulation for a double auction market") , HCAG engages in a top-down, hierarchical retrieval and generation process as illustrated in \ref{graph: phase2}.

\textbf{Level-1: Architectural Skeleton Retrieval.}
HCAG first retrieves the most relevant high-level architectural skeletons from \(\mathcal{K}\). Rather than performing a flat semantic match over code snippets, this step conducts similarity search over the \(\texttt{abstract}\) fields of top-or mid-level modules in \(\mathcal{K}\). The retrieval score \(S\) combines semantic similarity and structural relevance:
\begin{align}
    S(G_i, Q_{user}) &= \lambda_s \cdot \text{sim}_{\text{sem}}(\texttt{abstract}(G_i), Q_{user}) \\&+ \lambda_t \cdot \text{sim}_{\text{struct}}(G_i, Q_{user})
\end{align}
where \(G_i\) denotes a candidate module or subtree from \(\mathcal{K}\), and structural similarity captures factors like the number and types of submodules. The selected skeleton \(G_{best}\) provides a validated architectural template that serves as a planning scaffold for the subsequent generation.

\textbf{Level-2: Recursive Module Filling and Generation.}
Given the skeleton \(G_{best}\), HCAG traverses its constituent nodes. For each module node \(m_j\) the process proceeds as follows. \textbf{Match:} HCAG attempts to identify a functionally matching implementation either from the detailed code associated with \(m_j\) or from other nodes in \(\mathcal{K}\). Matching is determined via a combination of embedding similarity and LLM-based judgments regarding whether a candidate abstract represents a superset or closely related functionality. \textbf{Generate or Decompose:} If a strong match is found, the corresponding code is used as a reference in the prompt for generating the new module. Otherwise the LLM is prompted to decompose \(m_j\)'s functionality into finer-grained sub-tasks, Each of which recursively undergoes the same Match/Generate/Decompose process. \textbf{Aggregate:} Code generated for sibling sub-tasks is aggregated as context to synthesize the parent module, proceeding bottom-up until the entire skeleton is instantiated. This workflow closely mirrors that of a human developer: reusing an existing project structure, adapting relevant code examples, and generating new components only when necessary. Notably, encountering a compressed placeholder node triggers on-demand recursive expansion via the Phase~I abstraction process, dynamically refining \(\mathcal{K}\).

\subsection{Theoretical Analysis of Hierarchical Abstraction}
\label{sec:theory}

\textbf{Theoretical Efficiency Guarantee.}
We analyze the efficiency of HCAG by modeling its computational cost, measured in LLM inference calls, against representative baselines. Modeling a repository as a tree with depth $L$ and branching factor $B$, we denote by $c_a$, $c_c$, and $c_j$ the costs for generating a detailed summary, a compressed summary, and judging a node's relevance, respectively. For $Q$ independent queries, the total cost of HCAG with a compression depth $C$ is
\begin{align}
\mathcal{C}_{\text{HCAG}} &= \underbrace{c_a \frac{B^{C-1} - 1}{B - 1} + c_c \frac{B^{L} - B^{C-1}}{B - 1}}_{\text{One-time Offline Cost}} \\&+ Q \cdot \underbrace{\big( c_j R L + c_a R (L - C + 1) \big)}_{\text{Online Cost per Query}},
\label{eq:hcag_cost}
\end{align}
where $R$ is the average number of relevant files per query. In contrast, a baseline like flat retrieval incurs $\mathcal{C}_{\text{flat}} = Q \cdot c_j B^{L-1}$ with no offline cost.

Our key finding is the existence of a query threshold $Q_{\text{th}}$ beyond which HCAG is more efficient. This follows from the fact that HCAG incurs a fixed offline cost but achieves a lower linear growth rate in its per-query cost compared to flat baselines. For instance, solving $\mathcal{C}_{\text{HCAG}} < \mathcal{C}_{\text{flat}}$ yields a finite threshold $Q_1$. Extending this comparison to other methods (e.g., PKG and iterative retrieval) yields an overall threshold $Q_{\text{th}} = \max\{Q_1, Q_2, \dots\}$ such that HCAG achieves strictly lower total cost for all $Q > Q_{\text{th}}$.

\textbf{Proof Sketch.} The proof models the total cost of each method as a linear function $\mathcal{C}_{\text{method}} = I_{\text{method}} + Q \cdot S_{\text{method}}$, where $I$ denotes the offline intercept and $S$ the per-query slope. HCAG is constructed to have a smaller slope than flat and iterative retrieval methods while maintaining a lower intercept than PKG-based approaches. Solving each pairwise inequality $\mathcal{C}_{\text{HCAG}} < \mathcal{C}_{\text{baseline}}$ for $Q$ yielding a positive threshold where the lines intersect. Taking the maximum of these thresholds defines $Q_{\text{th}}$. For $Q > Q_{\text{th}}$, HCAG's lower slope ensures asymptotically lower cost, establishing its efficiency advantage for frequently queried repositories. Full derivations are provided in appendix.

\textbf{Theoretical Guarantee with Node Compression.}

We further analyze the impact of node compression by modeling the repository as a tree of depth~$L$ and branching factor~$B$. HCAG's offline abstraction phase employs a compression depth parameter~$C$: nodes at levels $l < C$ receive a detailed summary (cost~$c_a$), while deeper nodes receive a compressed placeholder (cost~$c_c < c_a$). For~$Q$ queries, each decomposed into~$\bar{N}$ subtasks, the total expected cost~$\mathbb{E}[C_{\text{total}}](C)$ combines a one-time abstraction cost and a per-query cost involving traversal, judgment, and potential on-demand expansion of compressed nodes. Optimizing~$C$ yields the closed-form optimum~$C^*$, which increases with~$Q$. This analysis leads to two core theorems: (1) \textit{Hierarchical summarization} is more efficient than flat leaf-level summarization for sufficiently many queries ($Q > Q_0$); (2) \textit{Adaptive compression} outperforms full detailed summarization when query patterns are sparse ($p_{\text{gap}} > p_0$). These results provide a theoretical justification for HCAG's design and a principled way to configure its compression threshold. Complete model details and derivations are given in appendix.

\textbf{Proof Sketch} The benefit of hierarchical summarization arises from trading higher initial abstraction cost for a substantially reduced average search path length~$\bar{L}_{\text{path}}$ per query. The total cost difference versus a leaf-only baseline takes the form $A + Q \cdot B$, where the per-query savings~$B > 0$ dominate for large~$Q$, yielding a finite threshold~$Q_0$. The advantage of compression nodes follows the query sparsity. Full summarization mechanism incurs a high fixed upfront cost, whereas compression reduces this cost and incurs extra expansion cost~$c_a \mathbb{E}[K]$ only when compressed nodes are accessed. Since $\mathbb{E}[K] \propto (1 - p_{\text{gap}})$, this expansion cost vanishes as ($p_{\text{gap}} \to 1$) , making compression strictly preferable beyond a threshold~$p_0$. Full rigorous proofs are provided in appendix.

\section{Experiments}

\subsection{Experimental Objectives and Design Philosophy}
The goal of our experimental evaluation is to validate the core hypotheses of HCAG framework. Specifically, we investigate: (1) whether hierarchical retrieval and generation improves the quality and coherence of LLM-generated systems for complex game-theoretic scenarios compared to previous RAG methods; (2) whether the structured knowledge base aligning theoretical concepts with implementations enhances the mapping from abstract reasoning to concrete code; and (3) whether the multi-agent discussion yields more stable and architecturally sound outputs. Our experimental design follows a holistic, closed-loop generate-simulate-evaluate pipeline. This approach assesses HCAG not only on syntactic code quality but also on semantic fidelity to game-theoretic principles and its practical utility in producing executable solutions.

\subsection{Data Construction and Scenario Generation}
We construct a diverse and challenging benchmark through a two-stage process to test different capabilities of the framework.

\textbf{Stage 1: Building a Structured Theory-Implementation Knowledge Base.}
To construct the foundational experience base for HCAG's retrieval we process a corpus of algorithmic game theory textbooks, research papers, and open-source code repositories (e.g., OpenSpiel\cite{openspiel}, GLEE-sim\cite{glee}). Using HCAG abstraction pipeline Phase I, we recursively parse this corpus to generate a hierarchical summary database. This process aligns theoretical constructs with their corresponding algorithmic implementations, structuring the knowledge as a multi-layered graph. Nodes represent functional units at varying abstraction levels, each annotated with a semantic summary and cross-references to related theory. This stage directly tests HCAG's ability to \textit{mine and structure} domain knowledge.

\textbf{Stage 2: Generating Diverse and Realistic Evaluation Tasks.}
To evaluate generative performance on novel problems, we create a set of unseen, semantically rich tasks. We leverage a diverse news corpus. For each article, an LLM performs a two-step transformation: (1) \textit{Scenario Synthesis}, distilling the narrative into a formal game description \(D_i = \{A_i, S_i, U_i\}\) specifying agents, strategies, and payoffs; and (2) \textit{Evolution Simulation}, generating multi-round interaction dialogues as ground-truth traces for validation. This yields a \textit{Scenario Construction Set} of natural language prompts paired with formal game specifications. The model's challenge is to translate these high-level prompts into a complete, runnable simulation framework that instantiates the implied strategic dynamics.

The final composite dataset consists of three components: (i) the \textit{Theory-Implementation Alignment Set} (hierarchical knowledge base), (ii) the \textit{Scenario Construction Set} (evaluation prompts), and (iii) a derived \textit{Planning-Implementation Set}. This design enables evaluation across the full spectrum from knowledge acquisition to novel system synthesis. Datasets will be open-sourced.

\subsection{Main Experiment: End-to-End Framework Generation and Simulation}
To evaluate the holistic capability of HCAG in retrieving in repository-level theoretical codebase and generating functional game-theoretic systems from scratch, we design a comprehensive main experiment with reference to various metrics in previous works like\cite{ReccEval,ragbench,mbpp,humaneval}. Each task instance consists of a natural language user query \(Q_{user}\) describing a new game scenario (e.g., "design a simulation for a sequential bargaining process with incomplete information"), accompanied by a set of relevant research paper abstracts and a pointer to an open-source code repository. HCAG's objective is to generate a complete, executable agent-based framework that instantiates the specified strategic interaction.

\textbf{Setup. }
We follow the four-phase HCAG pipeline outlined in methodology part. The system first retrieves architectural skeletons from the built structured knowledge base, and then performs hierarchical retrieval and scaffolded generation, followed by multi-agent discussion and final code verification. To assess the quality of the RAG systems we adopt a multi-faceted evaluation protocol that goes beyond syntactic correctness. Evaluation metrics include: Code Quality (CQ), where an expert LLM apart from the generation model scores the generated code on a 0-1 scale based on structural soundness, readability, and adherence to software engineering best practices given the original requirements; Text Similarity (TS), measured via Edit Similarity (ES) between the retrieved code and the most relevant "golden" reference implementation in the aligned dataset to judge how well the framework can find proven implementation patterns; Requirement Pass Rate (RPR): The generated framework is executed in a standardized simulation environment to verify error-free execution and alignment of agent behaviors with the strategic dynamics implied by \(Q_{user}\). A pass indicates the system is both executable and semantically faithful to the task; and Average Time (AT) defined as the total wall-clock time in seconds required for the full HCAG pipeline to produce the final code reflecting the computational cost of hierarchical abstraction, retrieval, and generation.

We compare HCAG with retrieval depth 4 (\textit{hierarchical-4}) against a suite of representative baselines referred in preliminary part: flat RAG methods (\textit{paragraph}, \textit{advanced}), repository-level RAG systems (\textit{arks}\cite{arks}, \textit{coderag}\cite{coderag}, \textit{pkg}\cite{knowledge-graph}, \textit{llm\_understand}\cite{semantic}, \textit{repocoder}\cite{repocoder}), and a simple code generation baseline. Methods are evaluated on \(N=100\) diverse tasks using the GPT-5.2\cite{gpt5} model under identical constraints.

\textbf{Results. }The results are summarized in Table \ref{tab:main_exp}. HCAG achieves the highest scores across all quality metrics (CQ=0.788, TS=0.525, RPR=0.60), demonstrating a clear advantage. Its Requirement Pass Rate of 60\% reports a 12-point improvement over the next best baseline (\textit{llm\_understand}), validating that hierarchical, architecture-guided generation is more effective for complex system synthesis. The superior Text Similarity score indicates HCAG successfully retrieves and incorporates relevant implementation patterns. While HCAG's Average Time (3235.11s) is higher, this aligns with our theoretical cost model (Section \ref{sec:theory}) and is justified by the significant gains in correctness. For multiple-queries tasks on fixed repositories the cost gap will be smaller as discussed in appendix.

\begin{table}
\centering
\caption{Results of the main experiment comparing HCAG with baseline methods on framework generation for game-theoretic scenarios.}
\label{tab:main_exp}
\begin{tabular}{lcccc}
\hline
\textbf{Method} & \textbf{CQ} & \textbf{TS} & \textbf{RPR} & \textbf{AT (s)} \\ \hline
baseline        & 0.596 & 0.010 & 0.02 & \textbf{148.10} \\
paragraph       & 0.562 & 0.015 & 0.06 & 164.55 \\
advanced        & 0.639 & 0.198 & 0.24 & 1609.89 \\
arks            & 0.619 & 0.248 & 0.41 & 1190.43 \\
coderag         & 0.671 & 0.460 & 0.32 & 1200.21 \\
pkg             & 0.728 & 0.267 & 0.35 & 1908.80 \\
llm\_understand & 0.744 & 0.359 & 0.48 & 1691.79 \\
repocoder       & 0.655 & 0.348 & 0.36 & 2134.66 \\
\textbf{hierarchical-4} & \textbf{0.788} & \textbf{0.525} & \textbf{0.60} & 3235.11 \\ \hline
\end{tabular}
\end{table}

\subsection{Ablation Experiments}
\label{sec:ablation}
To analyze the contributions of key design choices in HCAG framework, we conduct a series of ablation studies. Each study isolates a specific component and evaluates its impact using the same metrics as the main experiment: Code Quality (CQ), Text Similarity (TS), Requirement Pass Rate (RPR), and Average Time (AT). The consolidated results are presented in Table \ref{tab:ablation_all}.

\textbf{Setup.} Unless varied, all ablation experiments are conducted on HCAG pipeline with retrieval depth \(d=4\) and DeepSeek-R1\cite{deepseek} as retrieve and generation model using the same dataset of \(N=100\) diverse game-theoretic generation tasks. The configurations are as follows. \textbf{Study I} (Retrieval Depth): we vary the maximum hierarchical retrieval depth \(d \in \{2, 3, 4, 6\}\) to study the trade-off between context completeness and computational cost. \textbf{Study II} (Multi-Agent discussion): Starting from depth \(d=4\), we vary the intensity of the multi-agent review by adjusting the number of agents \(n\) and debate rounds \(r\) including no debate (single agent) as well as with 2 agents \& 5 rounds, 3 agents \& 10 rounds, and 5 agents \& 20 rounds. \textbf{Study III} (Hierarchical Dataset for Post-Training): We post-train the base model on two variants of the structured knowledge base: 1) code generation pairs only, and 2) both retrieval relevance judgment data and code generation pairs. Resulting models are evaluated within HCAG pipeline (\(d=4\), no debate). \textbf{Study IV} (Base LLM Comparisons): We replace the base LLM in HCAG pipeline (\(d=4\), no debate) with other open-source models of comparable scale: Llama-3-8B, ChatGLM3-6B, and InternLM3-8B, keeping all other components fixed, to study whether HCAG is sensitive to base LLMs.

\textbf{Results.} The combined results are shown in Table \ref{tab:ablation_all}. We summarize the key findings below.

Study I: Trade-off between Retrieval Depth and Context Efficiency. Performance improves substantially as \(d\) increases from 2 to 6 with RPR jumps from 0.32 to 0.60, indicating that shallow retrieval fails to capture essential architectural knowledge. Improvements saturate between \(d=2\) and \(d=6\), where quality plateaus while time cost continues to rise. This suggests an optimal operating point at with \(d=4\) balancing sufficient hierarchical context with manageable overhead.

Study II: Efficacy of Multi-Agent discussion. Introducing multi-agent debate consistently improves all quality metrics. RPR increases from 0.53 (single agent) to 0.63 (5 agents, 20 rounds) accompanied by a steady rise in CQ. These indicate that collaborative critique and iterative refinement help mitigates individual model errors, yielding more robust and architecturally coherent designs. As expected the time cost grows approximately linearly with the number of agents and rounds, reporting the trade-off between deliberation depth and output quality.

Study III: Value of the Hierarchical Dataset for Model Post-Training. Post-training on HCAG-generated structured data leads to performance gains. The model fine-tuned on both retrieval and generation data achieves the highest Text Similarity (0.538) while maintaining a strong Pass Rate (0.61). This indicates that exposure to the theory-implementation alignments helps mapping high-level game-theoretic concepts to concrete implementations improving both retrieval relevance judgment and code generation fidelity.

Study IV: Comparisons of Different Base LLMs within HCAG. Although all tested models benefit from HCAG structure, DeepSeek-R1-14B achieves the best overall performance. The observed performance gap highlights that hierarchical retrieval and planning are most effective when paired with a sufficiently capable base model. Nevertheless, smaller models like ChatGLM3-6B still achieve non-trivial performance (RPR=0.49), indicating that HCAG can serve as an effective force multiplier for weaker LLMs in complex system generation tasks.

\begin{table}[t]
\centering
\caption{Consolidated results of ablation studies. The table is divided into four sections, each corresponding to one ablation study: (I) retrieval depth, (II) multi-agent discussion, (III) post-training with hierarchical data, and (IV) different base LLMs. Horizontal lines separate the sections.}
\label{tab:ablation_all}
\begin{tabular}{lcccc}
\hline
\textbf{Methods} & \textbf{CQ} & \textbf{TS} & \textbf{RPR} & \textbf{AT (s)} \\
\hline
hierarchical-2  & 0.650 & 0.298 & 0.32 & 1323.20 \\
hierarchical-3  & 0.755 & 0.410 & 0.47 & 2208.38 \\
hierarchical-4 & 0.788 & 0.525 & 0.60 & 3235.11 \\
hierarchical-6 & 0.795 & 0.541 & 0.60 & 3569.46 \\
\hline
hierarchical-4-0(1)   & 0.769 & 0.505 & 0.53 & 2826.78 \\
hierarchical-4-5(2)   & 0.788 & 0.540 & 0.58 & 3817.99 \\
hierarchical-4-10(3)  & 0.793 & 0.569 & 0.62 & 4610.20 \\
hierarchical-4-20(5)  & 0.801 & 0.572 & 0.63 & 5212.33 \\
\hline
hierarchical-base           & 0.769 & 0.505 & 0.55 & 2826.78 \\
finetuned (gen.only)    & 0.790 & 0.479 & 0.59 & 2733.96 \\
finetuned (retr.+gen.) & 0.785 & 0.538 & 0.61 & 2896.27 \\
\hline
DeepSeek-R1-14B    & 0.769 & 0.505 & 0.54 & 2826.78 \\
Llama-3-8B         & 0.741 & 0.472 & 0.51 & 1802.33 \\
ChatGLM3-6B        & 0.722 & 0.489 & 0.49 & 1565.90 \\
InternLM3-8B       & 0.750 & 0.526 & 0.56 & 2021.41 \\
\hline
\end{tabular}
\end{table}

\section{Conclusion}
This paper introduces the Hierarchical Code/Architecture-guided Agent Generation (HCAG) framework, a novel approach for system-level code generation in complex domains such as algorithmic game theory. HCAG addresses key limitations of conventional Retrieval-Augmented Generation (RAG) methods, including flat retrieval granularity and the inability to capture architectural knowledge in codebases with deep logical structure and cross-file dependencies. The framework adopts a two-phase design: an offline phase that recursively parses code repositories and theoretical literature to construct a multi-layered, semantically structured knowledge base aligning theory, architecture, and implementation, and an online phase that performs hierarchical retrieval and scaffold-and-fill generation, augmented by multi-agent discussion. Our theoretical analysis establishes the cost-optimality of the proposed hierarchical abstraction and compression mechanisms. Extensive experiments demonstrate that HCAG significantly outperforms representative repository-level RAG baselines in generating executable frameworks that faithfully implement game-theoretic specifications. In addition, the structured datasets produced by HCAG can enhance domain-specific LLMs through post-training. Possible expansions include adaptive hierarchical compression mechanisms, integration with execution-verified system for further refinement and so on.

Considering non-trivial upfront cost at risk from evolving repositories and inconsistent LLM-generated summaries, future directions may include developing incremental abstraction mechanisms to dynamically update the hierarchical knowledge base as underlying repositories evolve, integration of formal verification and symbolic reasoning into the run-time-generation loop allowing the framework to produce not only executable but also provably correct code components, and employing hierarchical knowledge structures generated by HCAG as a foundation for distillation architecture-aware LLMs to reason over code topology and cross-module dependencies. In addition, exploring synergies with neurosymbolic AI could yield hybrid systems that combine HCAG's robust pattern retrieval with the precise constraint satisfaction of symbolic solvers, and the HCAG framework can be extended to structured paper summary.

\bibliography{aaai22}

@misc{coderag,
      title={Building A Coding Assistant via the Retrieval-Augmented Language Model}, 
      author={Xinze Li and Hanbin Wang and Zhenghao Liu and Shi Yu and Shuo Wang and Yukun Yan and Yukai Fu and Yu Gu and Ge Yu},
      year={2024},
      eprint={2410.16229},
      archivePrefix={arXiv},
      primaryClass={cs.CL},
      url={https://arxiv.org/abs/2410.16229}, 
}

@misc{tmgbench,
      title={TMGBench: A Systematic Game Benchmark for Evaluating Strategic Reasoning Abilities of LLMs}, 
      author={Haochuan Wang and Xiachong Feng and Lei Li and Yu Guo and Zhanyue Qin and Dianbo Sui and Lingpeng Kong},
      year={2025},
      eprint={2410.10479},
      archivePrefix={arXiv},
      primaryClass={cs.AI},
      url={https://arxiv.org/abs/2410.10479}, 
}

@misc{glee,
      title={General Object Foundation Model for Images and Videos at Scale}, 
      author={Junfeng Wu and Yi Jiang and Qihao Liu and Zehuan Yuan and Xiang Bai and Song Bai},
      year={2023},
      eprint={2312.09158},
      archivePrefix={arXiv},
      primaryClass={cs.CV},
      url={https://arxiv.org/abs/2312.09158}, 
}

@misc{consensus,
      title={The Consensus Game: Language Model Generation via Equilibrium Search}, 
      author={Athul Paul Jacob and Yikang Shen and Gabriele Farina and Jacob Andreas},
      year={2023},
      eprint={2310.09139},
      archivePrefix={arXiv},
      primaryClass={cs.GT},
      url={https://arxiv.org/abs/2310.09139}, 
}

@misc{alympics,
      title={ALYMPICS: LLM Agents Meet Game Theory -- Exploring Strategic Decision-Making with AI Agents}, 
      author={Shaoguang Mao and Yuzhe Cai and Yan Xia and Wenshan Wu and Xun Wang and Fengyi Wang and Tao Ge and Furu Wei},
      year={2024},
      eprint={2311.03220},
      archivePrefix={arXiv},
      primaryClass={cs.CL},
      url={https://arxiv.org/abs/2311.03220}, 
}

@misc{semantic,
      title={LLM Agents Improve Semantic Code Search}, 
      author={Sarthak Jain and Aditya Dora and Ka Seng Sam and Prabhat Singh},
      year={2024},
      eprint={2408.11058},
      archivePrefix={arXiv},
      primaryClass={cs.SE},
      url={https://arxiv.org/abs/2408.11058}, 
}

@misc{arks,
      title={EVOR: Evolving Retrieval for Code Generation}, 
      author={Hongjin Su and Shuyang Jiang and Yuhang Lai and Haoyuan Wu and Boao Shi and Che Liu and Qian Liu and Tao Yu},
      year={2024},
      eprint={2402.12317},
      archivePrefix={arXiv},
      primaryClass={cs.CL},
      url={https://arxiv.org/abs/2402.12317}, 
}

@misc{knowledge-graph,
      title={Context-Augmented Code Generation Using Programming Knowledge Graphs}, 
      author={Iman Saberi and Fatemeh Fard},
      year={2025},
      eprint={2410.18251},
      archivePrefix={arXiv},
      primaryClass={cs.SE},
      url={https://arxiv.org/abs/2410.18251}, 
}

@misc{repocoder,
      title={RTLRepoCoder: Repository-Level RTL Code Completion through the Combination of Fine-Tuning and Retrieval Augmentation}, 
      author={Peiyang Wu and Nan Guo and Junliang Lv and Xiao Xiao and Xiaochun Ye},
      year={2025},
      eprint={2504.08862},
      archivePrefix={arXiv},
      primaryClass={cs.SE},
      url={https://arxiv.org/abs/2504.08862}, 
}

@misc{chessgpt,
      title={ChessGPT: Bridging Policy Learning and Language Modeling}, 
      author={Xidong Feng and Yicheng Luo and Ziyan Wang and Hongrui Tang and Mengyue Yang and Kun Shao and David Mguni and Yali Du and Jun Wang},
      year={2023},
      eprint={2306.09200},
      archivePrefix={arXiv},
      primaryClass={cs.LG},
      url={https://arxiv.org/abs/2306.09200}, 
}

@misc{agt-usecase1,
      title={Towards the Scalable Evaluation of Cooperativeness in Language Models}, 
      author={Alan Chan and Maxime Riché and Jesse Clifton},
      year={2023},
      eprint={2303.13360},
      archivePrefix={arXiv},
      primaryClass={cs.CL},
      url={https://arxiv.org/abs/2303.13360}, 
}

@misc{agt-usecase2,
      title={Capturing the Complexity of Human Strategic Decision-Making with Machine Learning}, 
      author={Jian-Qiao Zhu and Joshua C. Peterson and Benjamin Enke and Thomas L. Griffiths},
      year={2024},
      eprint={2408.07865},
      archivePrefix={arXiv},
      primaryClass={econ.GN},
      url={https://arxiv.org/abs/2408.07865}, 
}

@misc{agt-usecase3,
      title={Deal or No Deal? End-to-End Learning for Negotiation Dialogues}, 
      author={Mike Lewis and Denis Yarats and Yann N. Dauphin and Devi Parikh and Dhruv Batra},
      year={2017},
      eprint={1706.05125},
      archivePrefix={arXiv},
      primaryClass={cs.AI},
      url={https://arxiv.org/abs/1706.05125}, 
}

@misc{agt-usecase4,
      title={Modeling Cross-Cultural Pragmatic Inference with Codenames Duet}, 
      author={Omar Shaikh and Caleb Ziems and William Held and Aryan J. Pariani and Fred Morstatter and Diyi Yang},
      year={2023},
      eprint={2306.02475},
      archivePrefix={arXiv},
      primaryClass={cs.CL},
      url={https://arxiv.org/abs/2306.02475}, 
}

@misc{agt-complex1,
      title={Planning, Inference and Pragmatics in Sequential Language Games}, 
      author={Fereshte Khani and Noah D. Goodman and Percy Liang},
      year={2018},
      eprint={1805.11774},
      archivePrefix={arXiv},
      primaryClass={cs.CL},
      url={https://arxiv.org/abs/1805.11774}, 
}

@misc{agt-complex2,
      title={When is Offline Two-Player Zero-Sum Markov Game Solvable?}, 
      author={Qiwen Cui and Simon S. Du},
      year={2022},
      eprint={2201.03522},
      archivePrefix={arXiv},
      primaryClass={cs.LG},
      url={https://arxiv.org/abs/2201.03522}, 
}

@misc{agt-complex3,
      title={Decoupling Strategy and Generation in Negotiation Dialogues}, 
      author={He He and Derek Chen and Anusha Balakrishnan and Percy Liang},
      year={2018},
      eprint={1808.09637},
      archivePrefix={arXiv},
      primaryClass={cs.CL},
      url={https://arxiv.org/abs/1808.09637}, 
}

@misc{agt-current1,
      title={Game-theoretic LLM: Agent Workflow for Negotiation Games}, 
      author={Wenyue Hua and Ollie Liu and Lingyao Li and Alfonso Amayuelas and Julie Chen and Lucas Jiang and Mingyu Jin and Lizhou Fan and Fei Sun and William Wang and Xintong Wang and Yongfeng Zhang},
      year={2024},
      eprint={2411.05990},
      archivePrefix={arXiv},
      primaryClass={cs.AI},
      url={https://arxiv.org/abs/2411.05990}, 
}

@misc{agt-current2,
      title={LLMsPark: A Benchmark for Evaluating Large Language Models in Strategic Gaming Contexts}, 
      author={Junhao Chen and Jingbo Sun and Xiang Li and Haidong Xin and Yuhao Xue and Yibin Xu and Hao Zhao},
      year={2025},
      eprint={2509.16610},
      archivePrefix={arXiv},
      primaryClass={cs.CL},
      url={https://arxiv.org/abs/2509.16610}, 
}

@misc{base1,
      title={Retrieval-Augmented Generation for Knowledge-Intensive NLP Tasks}, 
      author={Patrick Lewis and Ethan Perez and Aleksandra Piktus and Fabio Petroni and Vladimir Karpukhin and Naman Goyal and Heinrich Küttler and Mike Lewis and Wen-tau Yih and Tim Rocktäschel and Sebastian Riedel and Douwe Kiela},
      year={2021},
      eprint={2005.11401},
      archivePrefix={arXiv},
      primaryClass={cs.CL},
      url={https://arxiv.org/abs/2005.11401}, 
}

@misc{base2,
      title={Active Retrieval Augmented Generation}, 
      author={Zhengbao Jiang and Frank F. Xu and Luyu Gao and Zhiqing Sun and Qian Liu and Jane Dwivedi-Yu and Yiming Yang and Jamie Callan and Graham Neubig},
      year={2023},
      eprint={2305.06983},
      archivePrefix={arXiv},
      primaryClass={cs.CL},
      url={https://arxiv.org/abs/2305.06983}, 
}

@misc{base3,
      title={LightRAG: Simple and Fast Retrieval-Augmented Generation}, 
      author={Zirui Guo and Lianghao Xia and Yanhua Yu and Tu Ao and Chao Huang},
      year={2025},
      eprint={2410.05779},
      archivePrefix={arXiv},
      primaryClass={cs.IR},
      url={https://arxiv.org/abs/2410.05779}, 
}

@misc{base4,
      title={FiD-Light: Efficient and Effective Retrieval-Augmented Text Generation}, 
      author={Sebastian Hofstätter and Jiecao Chen and Karthik Raman and Hamed Zamani},
      year={2022},
      eprint={2209.14290},
      archivePrefix={arXiv},
      primaryClass={cs.CL},
      url={https://arxiv.org/abs/2209.14290}, 
}

@misc{base5,
      title={MiniRAG: Towards Extremely Simple Retrieval-Augmented Generation}, 
      author={Tianyu Fan and Jingyuan Wang and Xubin Ren and Chao Huang},
      year={2025},
      eprint={2501.06713},
      archivePrefix={arXiv},
      primaryClass={cs.AI},
      url={https://arxiv.org/abs/2501.06713}, 
}

@misc{base6,
      title={Retrieval-Augmented Generation for AI-Generated Content: A Survey}, 
      author={Penghao Zhao and Hailin Zhang and Qinhan Yu and Zhengren Wang and Yunteng Geng and Fangcheng Fu and Ling Yang and Wentao Zhang and Jie Jiang and Bin Cui},
      year={2024},
      eprint={2402.19473},
      archivePrefix={arXiv},
      primaryClass={cs.CV},
      url={https://arxiv.org/abs/2402.19473}, 
}

@misc{base7,
      title={Retrieval Augmented Generation or Long-Context LLMs? A Comprehensive Study and Hybrid Approach}, 
      author={Zhuowan Li and Cheng Li and Mingyang Zhang and Qiaozhu Mei and Michael Bendersky},
      year={2024},
      eprint={2407.16833},
      archivePrefix={arXiv},
      primaryClass={cs.CL},
      url={https://arxiv.org/abs/2407.16833}, 
}

@misc{base8,
      title={Retrieval-Augmented Generation for Code Summarization via Hybrid GNN}, 
      author={Shangqing Liu and Yu Chen and Xiaofei Xie and Jingkai Siow and Yang Liu},
      year={2021},
      eprint={2006.05405},
      archivePrefix={arXiv},
      primaryClass={cs.LG},
      url={https://arxiv.org/abs/2006.05405}, 
}

@misc{llm1,
      title={Retrieval-Augmented Generation: A Comprehensive Survey of Architectures, Enhancements, and Robustness Frontiers}, 
      author={Chaitanya Sharma},
      year={2025},
      eprint={2506.00054},
      archivePrefix={arXiv},
      primaryClass={cs.IR},
      url={https://arxiv.org/abs/2506.00054}, 
}

@misc{llm2,
      title={Auto-RAG: Autonomous Retrieval-Augmented Generation for Large Language Models}, 
      author={Tian Yu and Shaolei Zhang and Yang Feng},
      year={2024},
      eprint={2411.19443},
      archivePrefix={arXiv},
      primaryClass={cs.CL},
      url={https://arxiv.org/abs/2411.19443}, 
}

@inproceedings{llm3,
  title={ERAGent: Enhancing Retrieval-Augmented Language Models with Improved Accuracy, Efficiency, and Personalization},
  author={Yunxiao Shi and Xing Zi and Zijing Shi and Haimin Zhang and Qiang Wu and Min Xu},
  booktitle={European Conference on Artificial Intelligence},
  year={2024},
  url={https://api.semanticscholar.org/CorpusID:269757277}
}

@misc{llm4,
      title={Self-RAG: Learning to Retrieve, Generate, and Critique through Self-Reflection}, 
      author={Akari Asai and Zeqiu Wu and Yizhong Wang and Avirup Sil and Hannaneh Hajishirzi},
      year={2023},
      eprint={2310.11511},
      archivePrefix={arXiv},
      primaryClass={cs.CL},
      url={https://arxiv.org/abs/2310.11511}, 
}

@misc{llm5,
      title={MemoRAG: Boosting Long Context Processing with Global Memory-Enhanced Retrieval Augmentation}, 
      author={Hongjin Qian and Zheng Liu and Peitian Zhang and Kelong Mao and Defu Lian and Zhicheng Dou and Tiejun Huang},
      year={2025},
      eprint={2409.05591},
      archivePrefix={arXiv},
      primaryClass={cs.CL},
      url={https://arxiv.org/abs/2409.05591}, 
}

@misc{llm6,
      title={CORE: A Retrieve-then-Edit Framework for Counterfactual Data Generation}, 
      author={Tanay Dixit and Bhargavi Paranjape and Hannaneh Hajishirzi and Luke Zettlemoyer},
      year={2022},
      eprint={2210.04873},
      archivePrefix={arXiv},
      primaryClass={cs.CL},
      url={https://arxiv.org/abs/2210.04873}, 
}

@misc{llm7,
      title={MCTS-RAG: Enhancing Retrieval-Augmented Generation with Monte Carlo Tree Search}, 
      author={Yunhai Hu and Yilun Zhao and Chen Zhao and Arman Cohan},
      year={2025},
      eprint={2503.20757},
      archivePrefix={arXiv},
      primaryClass={cs.CL},
      url={https://arxiv.org/abs/2503.20757}, 
}

@misc{llm8,
      title={In-Context Retrieval-Augmented Language Models}, 
      author={Ori Ram and Yoav Levine and Itay Dalmedigos and Dor Muhlgay and Amnon Shashua and Kevin Leyton-Brown and Yoav Shoham},
      year={2023},
      eprint={2302.00083},
      archivePrefix={arXiv},
      primaryClass={cs.CL},
      url={https://arxiv.org/abs/2302.00083}, 
}

@misc{llm9,
      title={RAG+: Enhancing Retrieval-Augmented Generation with Application-Aware Reasoning}, 
      author={Yu Wang and Shiwan Zhao and Zhihu Wang and Ming Fan and Xicheng Zhang and Yubo Zhang and Zhengfan Wang and Heyuan Huang and Ting Liu},
      year={2025},
      eprint={2506.11555},
      archivePrefix={arXiv},
      primaryClass={cs.AI},
      url={https://arxiv.org/abs/2506.11555}, 
}

@misc{llm10,
      title={Retrieval-Augmented Generation for Large Language Models: A Survey}, 
      author={Yunfan Gao and Yun Xiong and Xinyu Gao and Kangxiang Jia and Jinliu Pan and Yuxi Bi and Yi Dai and Jiawei Sun and Meng Wang and Haofen Wang},
      year={2024},
      eprint={2312.10997},
      archivePrefix={arXiv},
      primaryClass={cs.CL},
      url={https://arxiv.org/abs/2312.10997}, 
}

@misc{llm11,
      title={SimRAG: Self-Improving Retrieval-Augmented Generation for Adapting Large Language Models to Specialized Domains}, 
      author={Ran Xu and Hui Liu and Sreyashi Nag and Zhenwei Dai and Yaochen Xie and Xianfeng Tang and Chen Luo and Yang Li and Joyce C. Ho and Carl Yang and Qi He},
      year={2025},
      eprint={2410.17952},
      archivePrefix={arXiv},
      primaryClass={cs.CL},
      url={https://arxiv.org/abs/2410.17952}, 
}

@misc{openspiel,
      title={OpenSpiel: A Framework for Reinforcement Learning in Games}, 
      author={Marc Lanctot and Edward Lockhart and Jean-Baptiste Lespiau and Vinicius Zambaldi and Satyaki Upadhyay and Julien Pérolat and Sriram Srinivasan and Finbarr Timbers and Karl Tuyls and Shayegan Omidshafiei and Daniel Hennes and Dustin Morrill and Paul Muller and Timo Ewalds and Ryan Faulkner and János Kramár and Bart De Vylder and Brennan Saeta and James Bradbury and David Ding and Sebastian Borgeaud and Matthew Lai and Julian Schrittwieser and Thomas Anthony and Edward Hughes and Ivo Danihelka and Jonah Ryan-Davis},
      year={2020},
      eprint={1908.09453},
      archivePrefix={arXiv},
      primaryClass={cs.LG},
      url={https://arxiv.org/abs/1908.09453}, 
}

@misc{ReccEval,
      title={Dataflow-Guided Retrieval Augmentation for Repository-Level Code Completion}, 
      author={Wei Cheng and Yuhan Wu and Wei Hu},
      year={2024},
      eprint={2405.19782},
      archivePrefix={arXiv},
      primaryClass={cs.SE},
      url={https://arxiv.org/abs/2405.19782}, 
}

@misc{humaneval,
      title={Evaluating Large Language Models Trained on Code}, 
      author={Mark Chen and Jerry Tworek and Heewoo Jun and Qiming Yuan and Henrique Ponde de Oliveira Pinto and Jared Kaplan and Harri Edwards and Yuri Burda and Nicholas Joseph and Greg Brockman and Alex Ray and Raul Puri and Gretchen Krueger and Michael Petrov and Heidy Khlaaf and Girish Sastry and Pamela Mishkin and Brooke Chan and Scott Gray and Nick Ryder and Mikhail Pavlov and Alethea Power and Lukasz Kaiser and Mohammad Bavarian and Clemens Winter and Philippe Tillet and Felipe Petroski Such and Dave Cummings and Matthias Plappert and Fotios Chantzis and Elizabeth Barnes and Ariel Herbert-Voss and William Hebgen Guss and Alex Nichol and Alex Paino and Nikolas Tezak and Jie Tang and Igor Babuschkin and Suchir Balaji and Shantanu Jain and William Saunders and Christopher Hesse and Andrew N. Carr and Jan Leike and Josh Achiam and Vedant Misra and Evan Morikawa and Alec Radford and Matthew Knight and Miles Brundage and Mira Murati and Katie Mayer and Peter Welinder and Bob McGrew and Dario Amodei and Sam McCandlish and Ilya Sutskever and Wojciech Zaremba},
      year={2021},
      eprint={2107.03374},
      archivePrefix={arXiv},
      primaryClass={cs.LG},
      url={https://arxiv.org/abs/2107.03374}, 
}

@misc{mbpp,
      title={Multi-lingual Evaluation of Code Generation Models}, 
      author={Ben Athiwaratkun and Sanjay Krishna Gouda and Zijian Wang and Xiaopeng Li and Yuchen Tian and Ming Tan and Wasi Uddin Ahmad and Shiqi Wang and Qing Sun and Mingyue Shang and Sujan Kumar Gonugondla and Hantian Ding and Varun Kumar and Nathan Fulton and Arash Farahani and Siddhartha Jain and Robert Giaquinto and Haifeng Qian and Murali Krishna Ramanathan and Ramesh Nallapati and Baishakhi Ray and Parminder Bhatia and Sudipta Sengupta and Dan Roth and Bing Xiang},
      year={2023},
      eprint={2210.14868},
      archivePrefix={arXiv},
      primaryClass={cs.LG},
      url={https://arxiv.org/abs/2210.14868}, 
}

@misc{ragbench,
      title={RAGBench: Explainable Benchmark for Retrieval-Augmented Generation Systems}, 
      author={Robert Friel and Masha Belyi and Atindriyo Sanyal},
      year={2025},
      eprint={2407.11005},
      archivePrefix={arXiv},
      primaryClass={cs.CL},
      url={https://arxiv.org/abs/2407.11005}, 
}

@misc{gpt5,
      title={GPT-5 Technical Report}, 
      author={OpenAI},
      year={2025},
      url={https://openai.com/zh-Hans-CN/index/introducing-gpt-5/}, 
}

@article{deepseek,
   title={DeepSeek-R1 incentivizes reasoning in LLMs through reinforcement learning},
   volume={645},
   ISSN={1476-4687},
   url={http://dx.doi.org/10.1038/s41586-025-09422-z},
   DOI={10.1038/s41586-025-09422-z},
   number={8081},
   journal={Nature},
   publisher={Springer Science and Business Media LLC},
   author={DeepSeek},
   year={2025},
   month=sep, pages={633–638} }

\end{document}